%% file: ms.tex
\documentclass[sigconf]{acmart}

\AtBeginDocument{%
  \providecommand\BibTeX{{%
    \normalfont B\kern-0.5em{\scshape i\kern-0.25em b}\kern-0.8em\TeX}}}

\usepackage{balance}
\usepackage{subfig}
\usepackage{xspace}
\usepackage{ulem}
\usepackage{booktabs}
\newcommand{\co}{clone\,\&\,own\xspace}

\newcommand{\vpbench}{vpbench\xspace}
\newcommand{\Vpbench}{Vpbench\xspace}
\usepackage[capitalize]{cleveref}
\crefname{section}{Sec.}{sections}
\Crefname{section}{Section}{Sections}

\newcommand\parhead[1]{\vspace{.00mm}\noindent\textbf{{#1.}}}

\begin{document}

\title[A  Generator Framework For Evolving Variant-Rich Software]{A Generator Framework For Evolving Variant-Rich Software}

\author{Christoph Derks}
\affiliation{%
 \institution{Ruhr University Bochum}
 \city{Bochum}
 \country{Germany}}

\author{Daniel Strüber}
\affiliation{
  \institution{Chalmers | University of Gothenburg}
\city{Gothenburg}
\country {Sweden}
}
\affiliation{%
  \institution{Radboud University}
  \city{Nijmegen}
  \country{Netherlands}}

\author{Thorsten Berger}
\affiliation{%
  \institution{Ruhr University Bochum}
  \city{Bochum}
  \country{Germany}}
\affiliation{
  \institution{Chalmers | University of Gothenburg}
\city{Gothenburg}
\country {Sweden}
}
\renewcommand{\shortauthors}{Derks, Strueber, Berger}

\definecolor{darkred}{rgb}{0.55, 0.0, 0.0}
\newcommand{\ds}[1]{\textcolor{violet}{\textbf{[DS:} #1]}}
\newcommand{\cd}[1]{\textcolor{darkred}{\textbf{[CD:} #1]}}
\newcommand{\tb}[1]{\textcolor{purple}{\textbf{[TB:} #1]}}
\newcommand{\edit}[1]{\textcolor{teal}{#1}}
\newcommand{\cross}[1]{\textcolor{olive}{\sout{#1}}}
\renewcommand{\cross}[1]{}
\renewcommand{\cd}[1]{}

\begin{abstract}
\looseness=-1
Evolving software is challenging, even more when it exists in many different variants. Such software evolves not only in time, but also in space---another dimension of complexity.
While evolution in space is supported by a variety of product-line and variability management tools, many of which originating from research, their level of evaluation varies significantly, which threatens their relevance for practitioners and future research. Many tools have only been evaluated on ad hoc datasets, minimal examples or available preprocessor-based product lines, missing the early \co phases and the re-engineering into configurable platforms---large parts of the actual evolution lifecycle of variant-rich systems.
Our long-term goal is to provide benchmarks to increase the maturity of evaluating such tools. However, providing manually curated benchmarks that cover the whole evolution lifecycle and that 
are detailed enough to serve as ground truths, is challenging.

\looseness=-1
We present the framework \textit{vpbench} to generates source-code histories of variant-rich systems.
Vpbench comprises several modular generators relying on evolution operators that systematically and automatically evolve real codebases and document the evolution in detail.
We provide simple and more advanced generators---e.g., relying on code transplantation techniques to obtain whole features from external, real-world projects. We define requirements and demonstrate how vpbench addresses them for the generated version histories, focusing on support for evolution in time and space, the generation of detailed meta-data about the evolution, also considering compileability and extensibility.
\end{abstract}



\maketitle

\input{introduction}

\input{relatedwork}
\input{framework}

\input{evaluation}

\input{prototype}

\input{casestudies}

\input{conclusion}

\bibliographystyle{ACM-Reference-Format}
\bibliography{ms}

\end{document}

%% file: introduction.tex
\section{Introduction}
\looseness=-1
Engineering variant-rich systems is challenging. Developers create variants to fulfill the needs of different markets or environments, taking care of large and customized sets of requirements from different stakeholders. %
Evolving such systems is even more challenging.
The field of software product lines addressed this need and established a huge portfolio of techniques to effectively create variant-rich-systems. %
The focus has long been on building software platforms proactively and then evolving it. However, many real-world product lines are adopted retroactively, typically evolving from variants realized using \co\,\cite{struber2019facing,dubinsky2013exploratory,rubin2013survey,Mahmood}. Recognizing this need, much research emerged over the last years on supporting \co development and the migration of those cloned variants into platforms, reflecting the typical evolution lifecycle of variant-rich systems. A portfolio of techniques for automatically locating features, managing and identifying clones, creating feature models, and re-engineering cloned source code into configurable platforms controlled by features, has been established.

\looseness=-1
A typical evolution lifecycle of a variant-rich system starts with \co, where developers clone whole repositories (representing a single variant), then modify the clone and evolve the clones separately. While being simple without causing much overhead at the beginning, when the system evolves
the number of variants increases, organizations are quickly facing huge maintenance problems.
A bug might appear in one variant and get propagated as part of the cloning process \cite{rattan2013software} and requires fixing it in each version separately.
In fact, the illustrated issue with bug fixing expands to a larger set of problems\,\cite{dubinsky2013exploratory}, including change propagation and keeping an overview over the variants.
Then, organizations need to re-engineer variants into configured platforms, which is again challenging and risky, easily disrupting an organization. Furthermore, evolving a platform as a complex system is the next challenge, as well as re-integrating variants that were opportunistically cloned out of the platform again\,\cite{krueger2020promotepl}.

\looseness=-1
Researchers developed many useful tools to support this development lifecycle%
\,\cite{struber2019facing}.
A core challenge is their evaluation. While benchmarks of full product lines exist,\cite{struber2019facing}, that is not the case for the whole evolution lifecycle. Such benchmarks would need to contain information detailing the evolution---e.g., when a feature was introduced or propagated among variants, especially when the feature is scattered. While of course open-source systems exist, augmenting them with this level of detail is challenging and laborious, and has only been done for smaller evolving systems\,\cite{ji2015maintaining}.
As such, the limited
availability of benchmarks for variant-rich systems is not surprising---they are just so difficult to create manually.
In fact, a recent study found only 3 out of 11 considered common evolution scenarios being fully supported by benchmarks\,\cite{struber2019facing}.
Even for the heavily researched area of \textit{feature identification and location}\,\cite{rubin2013survey}, many tools were not assessed through benchmarks\,\cite{dit2013feature}.
Often, researchers resorted to simple proofs of concept or created own datasets without making them public. %
A workaround was to study the evolution of optional features, which are easily identifiable in code (via preprocessor annotations).
While this strategy helps evaluating preprocessor-focused techniques (e.g., variability-aware type-checking of C code\,\cite{kastner2012type}), it misses the \co phases and mandatory features (which need to be identified to make them optional when re-engineering clones into a platform) and other use cases, such as supporting variant synchronization by detecting clones of features and propagating changes.
Also, preprocessor-annotated code is only available for certain languages (e.g., C).

\looseness=-1
We present \textit{vpbench}, an extensible framework for generating revision histories covering feature-oriented evolution scenarios commonly found in variant-rich systems.
It provides mechanisms to simulate the development of a variant-rich system: it evolves an initial codebase over time and automatically adds, removes and clones features, mutates assets and clones variants. Feature addition is realized using feature transplantation. %
The evolution is documented by meta-data, usable as ground truth for evaluations.
This synthetic version history can be used to benchmark tools that require such a version history as an input, such as feature identification and location\,\cite{dit2013feature,rubin2013survey}, re-engineering\,\cite{Assuncao:2017:fl} and code integration tools\,\cite{lillack2019intention}. Our design addresses requirements related to evolving systems in a feature-oriented way, whilst documenting it as meta-data, assuring compilability, and being language-independent and extensible.

\looseness=-1
We show the feasibility of our framework by instantiating it, specific to Java projects using the build tool Gradle and conduct two case studies to show that we can simulate system evolution for two different initial systems, including the transplantation of new functionality from third-party systems. 
Our framework aims to lift the maturity of current and future tools for evolving variant-rich systems by providing benchmarks for their evaluation and to be extensible for further advances to improve its generation capabilities.

\looseness=-1
We contribute:
\textbf{requirements} for a  generation framework; the actual \textbf{framework} providing components and instantiable concepts to generate version histories on top of external projects, from which features are transplanted%
; an \textbf{evaluation} comprising a \textbf{prototypical} instantiation with further implemented generators and operations (including feature transplantation) to show feasibility , and two \textbf{case studies} demonstrate its generation capabilities with respect to system evolution and performance; and
an \textbf{online appendix}\footnote{https://bitbucket.org/VPBench/vpbench} with our code and evaluation data.

%% file: relatedwork.tex
\section{Background and Related Work}
\label{sec:relatedwork}

\parhead{Benchmark generation}
\looseness=-1
To generate an effective benchmark and potentially reuse insights from other benchmark generators, we surveyed existing system generation techniques.

One line of work follows a \textit{generation-from-scratch} strategy that generates a system given some input parameters.
W\"agemann et al. \cite{waegemann2017gene} propose an iterated process of selecting and inserting programming patterns from a given library into an emerging program such that it ensures a designated program input to lead to the worst-case execution time, providing the ground-truth for the final benchmark. %
Further related work stems from the domain of software verification \cite{steffen2014property,steffen2014tailored,jasper2019rers} or explores the generation of feature models \cite{mendonca2009sat,segura2012betty}.
A second line of work follows a \textit{generation from initial systems} strategy, building on input artifacts that are modified in some way to generate systems for benchmarking purposes. 
Kashyap et al.\,\cite{kashyap2019bug} provide a technique to generate a diverse set of bug-induced software by inserting bugs into an existing system. They execute the system and inspect the created dynamic traces to identify insertion points, into which they insert bugs from a library of bug templates. They generate multiple variants of the target system differing by injecting one bug in each.
Furthermore, there are lines of research on generating benchmark models in the model-driven engineering domain\,\cite{szarnyas2018train,varro2018towards,nassar2020emf,wu2018ocl} and performance testing of concrete software solutions \cite{zhu2007mdabench,bui2007benchmark,weiss2013systematic}. %
Our approach fits this strategy, too.
\looseness=-1
A third line of work is on \textit{remakes of systems}, i.e., reproducing an input system in a different way.
Martinez et al.\,\cite{martinez2018eclipse} select a subset of features on plug-in level from existent Eclipse IDE variants and combine them into new, executable variants. This process can be configured with a selection strategy, guaranteeing to hold feature constraints.
Other work 
generates benchmarks for software verification tools \cite{jasper2019rers}, 
introduced remaking systems for computer architecture and compiler design \cite{vanertfelde2010benchmark}, JavaScript engines  \cite{richards2011javascript} and model-based diagnosis \cite{wang2010diagnostic}. %

\looseness=-1
Unfortunately, no technique creates version histories. %
While a few papers proposed iterative techniques\,\cite{jasper2019rers,steffen2017petri,varro2018towards,waegemann2017gene,segura2012betty,nassar2020emf}, all of them exclusively focus on the end result, intermediate steps are not part of the produced system. This contrasts version histories, where intermediate versions are necessary parts of the result.
Whereas the surveyed approaches only add concrete functionality using predefined pattern libraries \cite{kashyap2019bug,wang2010diagnostic,waegemann2017gene}, which need to be created and maintained, we utilize automated code transplantation to implant new features. 
On a final note, we found only a single technique that actively generates program variants within a system\,\cite{martinez2018eclipse}, and one paper where such behaviour could be argued for \cite{kashyap2019bug}. %

\looseness=-1
\parhead{Code transplantation}
A core part of software development is the addition of new features.%
We study the automation of this task and take up the idea of automated code transplantation.
\textmu Scalpel\,\cite{barr2015automated} extracts an annotated organ using %
program slicing and implants it at a user-specified insertion point. It uses genetic programming to reduce the required slice size and find a variable mapping between host and donor that successfully completes all test cases. Similar to \textmu Scalpel, CodeCarbonCopy\,\cite{sidiroglou2017codecarboncopy} requires the user to provide the organ and insertion point. It extracts the specified functionality using a compile-time dependency graph and inserts it at the given insertion point. A variable mapping is created using symbolic expressions to convert the data representations of host and donor. This limits the applicability of this approach to programs working on the same input type. A search-based way of adding new functionality was proposed by Lu et al.\,\cite{lu2018program}. 
They don't transplant a specific organ, but search a database of code snippets to fill holes in a draft program such that it satisfies the program specification, given by I/O-tests or finite automata.
Zhang and Kim present the tool Grafter \cite{zhang2017automated}, which enables test reuse between code clones. They use five transplantation rules, that guarantee compilability on termination, to map a piece of code onto its clone. The mapping is based on concrete types or structural equivalence and closeness in naming. Finally, PatchWeave\,\cite{shariffdeen2020automated} tackles the Patch Transplantation Problem on two similar programs. They utilize version history information to extract the patch and find an insertion point in the target by identifying a divergent point in the version history of the fixed system and relating it to the erroneous system.

\section{Requirements}
\label{sec:requirements}
\looseness=-1
Our goal is to design and prototypically instantiate a benchmark generation framework capable of simulating %
the evolution of variant-rich software. It should take features from an existing project and generate a version history
simulating an evolution by automatically applying evolution operations while documenting the changes in detail. The long-term goal is to generate version histories that can be used in their entirety or partially for benchmarking software evolution techniques---the second important problem to be addressed in the future, but which requires a separate study with another methodology.
We consider the following requirements:

\looseness=-1
\parhead{Feature-oriented Evolution}
We rely on the assumption that systems are developed to some extent in terms of features, which are added, reused, removed, and so on. This assumption is reasonable to make, since developers have features in mind when implementing systems. Features are units of functionality, communication, and planning\,\cite{berger2015feature}.
However, developers usually do not make features (which can be cross-cutting) explicit in code, since recording them is not needed short-term. Of course, not all changes belong to features, but constitute changes to assets.
Unfortunately, the feature-oriented changes are not visible in the evolution history, since developers do not record features. 
In the evolution lifecycle, developers then typically need to recover this information (e.g., feature locations) to evolve or reuse (e.g., clone) features, or to re-engineer clones into a configurable platform\,\cite{Assuncao:2017:fl,jacob2020apogamesmigration}, making features optional and introducing configuration mechanisms. All these tasks are laborious and error-prone and call for tool support, which typically focuses on recovering such information and performing exact and precise re-engineering or refactoring tasks. However, current and future tools are hard to evaluate due to the general lack of benchmarks resembling real-world systems with a documented feature-oriented evolution that can serve as a ground truth.

So, in summary, the framework needs to provide mechanisms to evolve a system driven by feature changes (e.g., features added, cloned or removed). The features should be recorded in a feature model together with their locations in code.

\looseness=-1
\parhead{Meta-data}
A key requirement
is the exposure of exactly the kind of meta-data that is typically not recorded in practice, but needed to evaluate evolution techniques. Specifically, we need: (1) Feature locations, (2) clone traces between variants, and (3) information detailing the intentions of developers behind changes, related to implementing, maintaining and evolving features. %
For instance, when removing a feature, one might typically remove the feature-specific implementation assets as well. Recorded meta-data should clearly indicate this relation, i.e., the removal of implementation assets is a part of the high-level feature remove operation.

\looseness=-1
\parhead{Compilability} %
For a reasonable form of quality assurance for the generated artifacts, we demand compilability. The framework should provide a mechanism to ensure that every version of the generated software is at least compilable. Stronger guarantees, such as executability, are subject to future work, but very hard to achieve.

\parhead{Extensibility} 
The framework should be extensible with more code-manipulation techniques to simulate evolution (e.g., other types of changes), %
different algorithms for manipulating code (to account for different code characteristics one wants to generate), or to  generate code for different programming languages.

\parhead{Language-independence}
Finally, the framework should make no assumption regarding specific programming languages. 

%% file: framework.tex
\section{Generation Framework}
\label{sec:framework}

\label{sec:framework:ov}

\begin{figure}
	\centering
	\includegraphics[width=\linewidth]{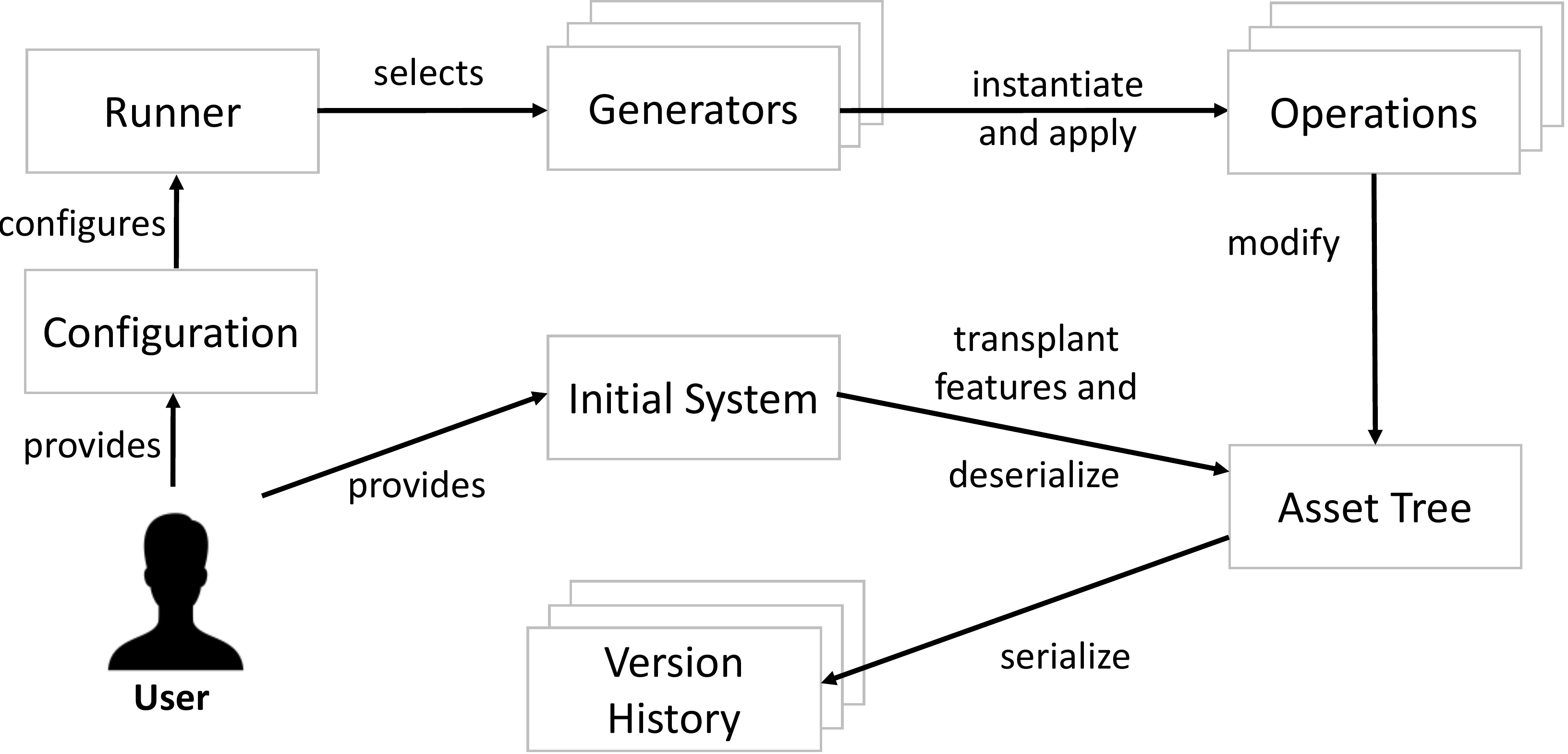}
	\vspace{-.5cm}
	\caption{Framework overview}
	\label{fig:main}
	\vspace{-.4cm}
\end{figure}

\looseness=-1
\Cref{fig:main} illustrates the framework and its interactions with the environment.
From the user perspective, it takes two inputs---a \textit{configuration} and an \textit{initial system}---and outputs a \textit{version history}.
The initial system consists of a codebase, i.e., the first revision, and a set of external projects to transplant features from. %
We now briefly introduce our main design decisions based on the requirements and the main components of our framework, each of which is is described in further details below.

\looseness=-1
\parhead{Components}
The framework provides extensible components to modify an abstract representation of the generated system evolution. Each generated revision (a codebase with folders and files) is internally represented as an \textit{asset tree}, a special abstract-syntax-tree-like format inspired by \textit{feature structure trees}\,\cite{apel.ea:2009:featurehouse} and borrowed from another framework called virtual platform\,\cite{Mahmood} for managing variant-rich systems. This asset tree is modified by dedicated \textit{operations}, extended from the virtual platform \cite{Mahmood}, which abstractly represent atomic evolutionary changes, some of which being feature-oriented to address the requirement ``feature-oriented evolution.''

\looseness=-1
Five operations are prescribed and provided as implementations or interfaces---to the extent the framework is still independent of the target programming language %
and other technology (e.g., build systems), addressing the requirement ``language independence.'' These operations are instantiated and applied by \textit{generators}, so these generate the actual changes on the asset tree, to serialize new revisions. Generators are executed by the coordinating \textit{runner}. It iteratively wraps operations in a transaction, thereby checking for the compilability requirements after application of operations 
to avoid faulty changes (at least with respect to compilation).

\parhead{Meta-Data}
The operators record \textit{meta-data}, which specifies their parametrization and their nested sub-operators.
After applying an operator, the system (i.e., the current version of the asset tree) is serialized, creating a snapshot of the system at a point in time. Through iteration (handled by the runner), we generate a version history of simulated changes to the user-provided initial system.

\looseness=-1
\parhead{Generators}
A generator is specialized to create some kind of change (i.e., operation type), on the evolving system. It simulates developers by generating suitable operation parameterizations that modify the asset tree.
The framework allows defining new operations by specifying modifications to the asset tree. %

\looseness=-1
\parhead{Framework Instantiation}
The framework facilities implementing common changes to variant-rich systems. It already provides programming-language-and technology-independent operators for simple changes (\texttt{Remove\-Feature}, \texttt{ChangeAsset}, \texttt{CloneRepository}).
More sophisticated ones, such as feature addition by code transplantation and feature cloning require more specialized implementations (e.g., combining simpler operations, or adding maintenance steps) and are provided in our prototype instantiation (\cref{sec:implementation}), but respective interfaces are part of the framework already.

\subsection{Revision History Representation}
\looseness=-1
An asset tree abstractly represents a variant-rich system as a tree with node types of different granularity, starting from the repository level via folders to files, and the sub-file level (e.g., methods, code blocks). It only keeps structure to the extent necessary to realize operations, but is otherwise almost fully language-independent.
Assets can map to features, which are stored inside of feature models that are associated with elements of the tree. The system is split in different repositories (which represent cloned system variants), all located beneath a synthetic root node. The asset tree contains all information---structure and node content (plain source code) to serialize it as the generated codebase in multiple revisions.

\looseness=-1
Our framework allows transplanting features from external projects. It models these by a project structure, defined by filepath, name, a folder for production and test code each, and optional subprojects, as supported by some build tools (e.g., Gradle).
Features of the same project can be added to multiple repositories. To this end, we store in which repositories a local project version is included and also include a list of available testcases in the project (explained shortly).

\subsection{Runner}
The runner operates as specified in \cref{sec:framework:ov}. It plays a coordinating role in the generation process, iteratively delegating change generation to generators and applying the generated changes to the system. To address the compilation requirement, we wrap concrete changes into transactions, that check the compilability of the changed system to filter faulty changes and discard them without applying them. The runner can be parametrized using six configuration options: (1.) a maximum amount of generation iterations and (2.) an optional termination condition on the evolving system (e.g., a certain number of features is included in the system). Generation terminates after the maximum number of iterations is reached or once the termination condition is fulfilled, whichever occurs first. (3.) The user defines the to-be-used set of generators and (4.) a static probability distribution, guiding the runner's selection of generators by assigning a probability for selection to each. Should no distribution be provided by the user, our implementation assumes a uniform distribution. Since generators are realized by stochastic processes, that might lead to invalid operations, i.e., operations leading to a non-compiling system, the user additionally defines (5.) a maximum number of retries one specific generator has to generate a valid operation, before moving to the next iteration and querying the next generator. Finally the user provides (6.) a concrete mechanism for checking system compilability (compilation checker), e.g., a specific build tool, to be used inside transactions. The runner implementation is part of the framework.

\subsection{Operations}\label{sec:framework:ops}
\looseness=-1
Operations specify blueprints of changes documented in meta-data and applied on the asset tree. The existing set of operations can be extended by describing how the asset tree gets changed. %
As part of our framework for evolving variant-rich software automatically, we provide five conceptual operators, which are inspired by a simulation study 
of a clone-based product line \cite{ji2015maintaining}: \textit{adding new assets as a new feature}%
, \textit{removing or disabling a feature}%
, \textit{cloning a project}%
, \textit{propagating a feature}%
(which we call \textit{cloning}), and \textit{evolving annotated assets}%
.  We now describe which of our operations realize these changes. %

\looseness=-1
\parhead{Remove Feature} 
Features can be removed again for a codebase. %
A feature gets selected and is removed from its corresponding feature model, including its subfeatures. As part of the process, all assets that are mapped to the selected feature only, are removed as well.

\looseness=-1
\parhead{Mutate Asset}
The content of a selected asset is modified in some form to simulate changes that developers might perform on implementation assets. We provide three simple mutation operators for adding, replacing and removing single lines of code. %

\looseness=-1
\parhead{Transplant Feature} %
Adding features is one of the most natural ways to evolve software. %
Adding new functionality to the system is much more complicated than the previous two operations. While work exists that automatically creates new functionality\,\cite{harman2014babel}, it requires defining testcases and ideally further guidance information. 

Instead of generating new features, our framework facilities feature transplantation\,\cite{barr2015automated,sidiroglou2017codecarboncopy,lu2018program,zhang2017automated,shariffdeen2020automated} from existing projects. This requires two inputs: the feature to transplant and where to insert it. However, it poses the following three problems.

\looseness=-1
\textit{Problem 1: What is a transplantable feature?} 
We approximate transplantable features using testcases. Similar to previous work\,\cite{li2017fhistorian} we assume test cases to call features to test their functionality. A feature for transplantation is identified by a testcase in an external system with the actual feature being the unit under test. %

\looseness=-1
\textit{Problem 2: How to extract a transplantable feature?}
We need to handle forward and backward dependencies\,\cite{barr2015automated}, i.e., the feature itself plus the code it calls as part of its execution and the code that prepares the execution environment for the feature, i.e., the vein. The former can be achieved by slicing the donor project down to the features dependencies. The latter is already provided by the testcase we use for feature identification, as testcases build an execution environment for their unit under test. One notable characteristic of our approach is that required assets, that are already part of the asset tree due to previous transplantation processes, are cloned from other repositories to the target repository. This maintains a sense of continuity and imitates a \co approach.

\looseness=-1
\textit{Problem 3: How to insert the extracted feature into the evolving system?}
The final hurdle is to integrate the new functionality in a sensible way. This typically requires finding a suitable variable mapping between host and donor. 
In our case this is simplified by the fact that the goal of feature transplantation is only to add some functionality.
Given an insertion point, i.e., a parent asset and insertion index, we add the testcase as a new asset at the defined position. This provides the evolving system with the necessary execution environment to execute the feature at this position in the program. The project slice we extracted in Problem 2 is added beneath the repository asset as a separate directory. Further transplantation processes that share dependencies are then integrated into previous project slices. %

\looseness=-1
\parhead{Clone Variant}
Typical variant-rich system evolution begins with \co. Existing variants are cloned and developed independently. %
The operation copies a selected variant and adds the clone to the asset tree as a sibling asset.

\looseness=-1
\parhead{Clone Feature}
A feature is cloned to another repository and added beneath a select feature in the target feature model. Cloning a feature requires cloning and integrating the feature implementing assets with the already present assets in the target. Depending on whether an asset is already contained in the target or not, we have to solve two different problems: (1.) The asset is already contained in the target, but potentially in a different version. In this case, we simply maintain the target version %
, though other behavior, e.g., taking into account version history information is possible, too. (2.) The asset is not contained in the target, but needs to be integrated \cite{lillack2019intention} with its siblings in the target, that might not exist beneath its parent in the source. This process is typically difficult to solve automatically, as code can not only be integrated in multiple ways, but in multiple dimensions: in variation points and in ordering.

\subsection{Generators}
\looseness=-1
Generators connect the simulation-coordinating runner with the operations. They generate instantiate operations with suitable parameterizations to actually change the asset tree. %
So, they simulate a developer in a two-step pipeline together with the runner. The runner selects a kind of change to be applied on the system by selecting a generator, which creates the concrete change, e.g., by selecting elements in the asset tree to apply it upon. The selection of parameters can be realized using stochastic processes. In line with the extensibility of operations, the framework can be extended with new generators. To this end, we provide implementation skeletons. %

\subsection{Meta-Data}\label{sec:framework:md}
\looseness=-1
To provide valuable ground-truths for different types of problems, our framework provides three types of meta-data as part of our simulated software evolution. We record feature locations using the asset-to-feature mappings, stored in the asset tree. An important part of variant-rich system evolution is cloning. We store clone traces, when elements inside the asset tree are cloned as part of applied operations. 
Additionally, we store meta-data on the applied operations itself.
The applied evolution patterns are recorded by storing the sequence of operations together with their parametrizations.

\looseness=-1
Key is a unique referencing of each element targeted by an operation. We solved this as follows: Filesystem-assets, i.e., repositories, folders and files, are uniquely addressed with their filepath. The same does not hold for codelevel-assets, e.g., classes or code blocks, as for example code blocks typically do not have a name. We reference codelevel-assets using a split path: the first half is the path of the containing filesystem-asset, the second half is the sequence of indices of the child-assets one has to pass through recursively to get from the containing filesystem-asset to the parametrized asset (index path)%
. Feature models are referenced by the assets containing them and features are identified by their feature model and their least-partially-qualified path (LPQ) therein\,\cite{schwarz2020common}. While these references are unique, they require a specific asset tree version to resolve them correctly. This recording also allows replying the evolutions.
Some operations work with elements outside the system's asset tree, e.g., when adding new code to the system. In these cases we store a representation of the entire external element, detailing its hierarchical structure and content. As part of recording operations, our framework also allows to recursively store suboperations with their parametrizations inside the calling operation%
. This realizes mapping low-level changes to high-level evolution intentions.

%% file: evaluation.tex
\section{Evaluation methodology}\label{sec:neweval}
We evaluate \vpbench by answering four questions:

\textbf{RQ1}. \textit{Is our framework realizable?} 
Implementing our framework poses significant technical challenges.
Operations and generators need to be a implemented in a way that supports the evolution of a real software project, in the context of its used programming language and build ecosystem.
The asset tree needs to be maintained in a way that supports its consistent modification and use for generation of new versions.
The most complicated of our operations is addressed in the next research question.

\textbf{RQ2}. \textit{Is feature transplantation realizable?} 
Feature transplantation is the most complex of our operations, raising three problems of identifying a transplantable feature, extracting it, and adding it to the evolving system in a useful way. 

\textbf{RQ3}. \textit{Can our framework implementation automatically evolve variant-rich systems?} 
We are interested in studying how the generated version histories look like.
Are they useful for simulating the evolution process of variant-rich software systems?

\textbf{RQ4}. \textit{What is the runtime performance of our framework implementation?} 
Since we might be dealing with complex projects and technologies, the runtime performance of our framework could potentially be a bottleneck for its potential applications.

To provide answers to these questions, we perform our evaluation in two steps: a prototype implementation and two case studies.
We now present both steps, together with the applied methodology.

\subsection{Prototype (RQ1+2)}
\looseness=-1
We show the feasibility of \vpbench by 
instantiating the framework in an implementation in Scala.
We implement the runner, a set of seven generators for creating the conceptual operations we introduced and provide these operations. Our implementation is specialised to work with the build tool Gradle for compilation and dependency management, i.e., both the evolving system and the external projects are Gradle projects. However, a large portion of our implementation can be reused to support other tools, too. 
Addressing RQ1, this implementation shows the feasibility of our framework conceptualization and its potential to generate configurable version histories.
Addressing RQ2, it contains an implementation of feature transplantation with a solution to the three outlined challenges of transplantable feature identification, extraction and addition.

\subsection{Case studies (RQ3+4)}
 To evaluate our framework's capability to simulate system evolution we empirically, conduct two case studies using our prototype, evolving a toy example system and a medium-sized system cloned from GitHub. To address RQ3, we present statistics from our simulation executions,  discussing system evolution in variability and size. To address RQ4, we examine the execution performance.

\looseness=-1
We generate version histories for two different initial systems over 500 iterations using different parameterizations. The selected initial systems are a small calculator example with 62 LoC and an open-source json-parser for Java\footnote{https://github.com/stleary/JSON-java} with 11,837 lines of code (LoC). For the latter we cloned the repository and applied some preprocessing, including adding two common repositories for dependency resolution to the projects' buildfile, to account for a current limitation (retrieving transitive Gradle dependencies that require retrieving and building third-party repositories).

\looseness=-1
We evolve these initial systems with three probability distributions. %
In all, we set a selection probability of $p = 0.01$ to both cloning generators due to scalability issues in memory consumption and runtime (cf. \cref{sec:eval:rq2}) and split up the remaining probability between the remaining five generators in the following way: a uniform distribution over the generators (\textit{Uniform Generators}), a uniform distribution over the remaining three types of conceptual changes (\textit{Uniform Operations}), that is, adding/removing a feature and changing an asset (probability is split uniformly for adding, deleting and replacing lines), and a distribution, that is expected to generate a growing system. The latter selects each line-changing-operation with a probability of $p = 0.2$, adds a new feature with $p = 0.29$, and removes a feature with $p = 0.09$ (\textit{Growing System}).

\looseness=-1
On selection, our stochastic generators have 50 attempts to generate a compilable change before aborting. The three line-changing operations discard ineffective changes with a probability of $p = 0.5$. We use two different donors for adding new features: the Structurizr client library\footnote{https://github.com/structurizr/java} and the HPC inter-thread messaging library LMAX Disruptor\footnote{https://github.com/LMAX-Exchange/disruptor}. We remove a task from the structurizr build scripts to allow us to generate the donor's jar files and compile the testcases as well as to generate the compilation mappings required for our transplantation implementation. We delete all multi-line comments from the disruptor library. 
The experimented were performed on a machine with an 3.6 GHz Intel Core i7-4790 processor and 8 GB RAM.

%% file: prototype.tex
\section{Prototype}
\label{sec:implementation}
\looseness=-1
We report on the results from our prototype implementation, addressing RQ1 (on our framework's realizability in a prototype) and RQ2 (on the realizability of feature transplantation).

\subsection{Basic Prototype (RQ1)}

\looseness=-1
	We implemented the five operations introduced in \cref{sec:framework} using seven generators.
	Below, we present the output of our technique as well as our operators together with the generators that implement them (with the exception of feature transplantation, which we present in the next subsection).

\looseness=-1
\parhead{Generation output}
The system we evolve consists of multiple variants, located beneath the synthetic root node. Each variant adheres to a Gradle multi-project build. It contains the initial system as a root project and project slices of the external projects as subprojects, as functionality gets added to the variant. Utilizing Gradle has benefits as it manages dependencies and should help to create more realistic systems due to its widespread use in practice, thus providing a large set of transplantable functionalities. We incorporate Gradle as a build tool to be used, when checking for compilability. %

\parhead{Remove Feature} We reuse the existing VP operation \texttt{RemoveFeature}. The generator selects the feature for removal using a uniform distribution over all features in the system.

\looseness=-1
\parhead{Mutate Asset} This operation extends the VP operation \texttt{ChangeAsset} to change the content of a specified asset. We provide three different generators, each performing a specific mutation operation, inspired by three out of nine program transformations proposed by Baudry et al.\,\cite{baudry2014tailored}, i.e., \textit{add-Random}, \textit{replace-Random} and \textit{delete}. 
Compared to the the original implementation, we implement these mutations on the line level, rather than the statement level. This allows the generators to work for any programming language. All three generators select a random asset and a random line $l_1$ to mutate. The generators \textit{add-Random} and \textit{replace-Random} select a second line $l_2$ from the same containing folder and insert $l_2$ before or instead of $l_1$. \textit{delete} simply removes $l_1$ from the system. We add optional sensibility checks, that discard some common ineffective changes, e.g. addition of an empty line, with a parametrized probability.

\looseness=-1
\parhead{Clone Variant} This is implemented using an extended version of the VP operator \texttt{CloneRepository}, which performs an additional maintenance step to update the external projects with information on the new variant that might include it%
. The generator invokes the operation on a repository selected according to a uniform distribution and creates a name for the new variant. Operation and generation are independent of programming language and build tool.

\parhead{Clone Feature} This operation is implemented as specified using an extended version of VP's \texttt{CloneFeature} operator to perform the automatic integration as described in \cref{sec:framework}. Similar to the transplantation operator, cloning features requires an extra step for dependency management, as new projects might be introduced in the target variant. We declare the newly added local project versions%
 and update the main evolving system's build file to define new dependencies. The dependency management step is first applied on the file system and only then converted to operations on the asset tree, similar to the feature transplanting operator. %
To be able to uniquely identify corresponding elements between source and target variant, we limit the applicability of this feature cloning process to repositories, that originated from each other (only from source to target).
This restriction allows us to use VP's clone traces to map elements that originated from each other between both variants. 
The generator thus selects a random feature that only exists in the source variant between two variants that originated from each other and invokes the operation.

\subsection{Feature Transplantation (RQ2)} 

We now discuss feature transplantation, as the most complex implemented operator, addressing the three problems of identifying, extracting and adding transplantable features.

\looseness=-1
\parhead{Transplant Feature} 
We create a new operation that takes as input a test case in an external project and an insertion point in the asset tree. We extract the feature for transplantation by differentiating between two types of dependencies: in-file and out-of-file dependencies. In-file dependencies can be elements such as statements executed in the constructor, import statements, attributes defined by the test class or local functions, that are called by the test case. For now, our implementation only supports the extraction of modular test cases (i.e., those that only require its file's import statements as in-file dependencies). These are extracted together with the test case itself using srcML \cite{maletic2002source}. srcML converts input source code into an XML representation, allowing for running queries on the program structure. Out-of-file dependencies are obtained using the Java dependency analyzer jdeps. As jdeps returns class-level dependencies based on class files, this results in a superset of dependencies. Note that our current implementation does not support dependencies on other files, that are located in the Gradle test source set. The hereby extracted class-level dependencies are mapped to their defining source-files and added to the organ. 

\looseness=-1
The in-file dependencies, i.e., test case and import-statements, are preprocessed (surrounded with a try-catch-block and receive an extra import for the test case's package respectively) and added as new code blocks at the insertion point to the main evolving system. All out-of-file dependencies are added as project slices from their original external projects. Transplanting only the the source code is not sufficient, we need to take care of its dependencies as well. To this end, we include the required external projects' build files in the project slice. As build files can grow arbitrarily complex, potentially inducing further problems, we apply an automated preprocessing step that parses, simplifies and adapts the original build file to be usable in our evolving system. Finally we update the main evolving system's build file to depend on the local project version that provided the transplanted test case. While our implementation for handling Gradle build files has some shortcomings, e.g., not automatically updating repositories or supporting only parts of the syntax, it works often enough, to show the feasibility of this approach.

\looseness=-1
These changes are first applied on the file system, checked for compilability and then converted to operations on the asset tree in the following manner: %
\textit{(1)} Testcase and import statements are added as new assets beneath their implantation point, \textit{(2)} other transplanted Java files are added as new assets, whilst storing their original file path as meta-data, or cloned from other repositories if possible, \textit{(3)} main project's Gradle files' assets are updated, \textit{(4.)} adapted build files of external projects are added as new assets, \textit{(5.)} test case, import statements and all recursive code dependencies are mapped to a new feature with the test case's name, which is added to the repository's feature model.

To prepare for feature transplantation, we identify test cases for transplantation during setup using a technique by Mukelabai et al.\,\cite{mukelabai2021semi} to identify annotated test cases, e.g. \texttt{@Test}, using srcML. %
The generator provides two types of input: a test case and an insertion point, both selected according to uniform distributions. Features are only inserted into methods in the current version of the initial system and not in local versions of external projects. This increases the likelihood of success by removing the possibility of introducing dependency circles. In case of an unsuccessful transplantation attempt, the test case is discarded and no implantation at a different location is attempted again.
While our implementation is Java- and Gradle-dependent, we are confident that this technique can similarly be implemented for other languages and build tools.

%% file: casestudies.tex
\section{Case Studies}\label{sec:evaluation}
\looseness=-1
We report on the results of our experimental evaluation on two cases, in which we generated version histories based on our prototype and two initial systems to answer RQ3 (on automated evolution of variant-research systems) and RQ4 (on runtime performance).

\subsection{Simulated Evolution (RQ3)}

We address whether our prototype can generate version histories showcasing evolution, i.e., if we can generate versions that evolve in variability and size over the course of a generated version history.

\begin{figure}
\centering
\vspace{-.6cm}
\subfloat[Calculator]{\includegraphics[trim=0cm 0cm 0.3cm 0.5cm, clip , width=0.5\textwidth]{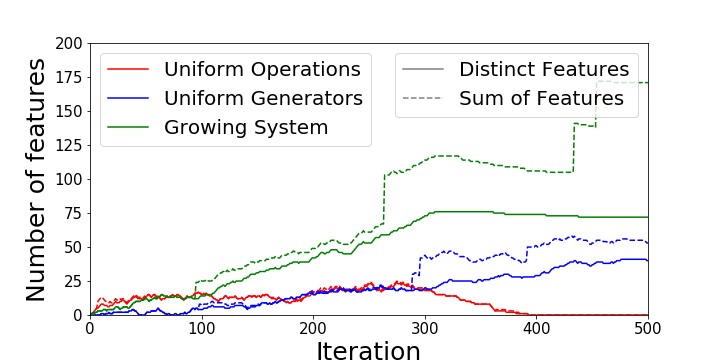}\label{fig:calcNumFeat}}
\hfill
\vspace{-.4cm}
\subfloat[JSON-java]{\includegraphics[trim=0cm 0cm 0.3cm 0.5cm, clip , width=0.5\textwidth]{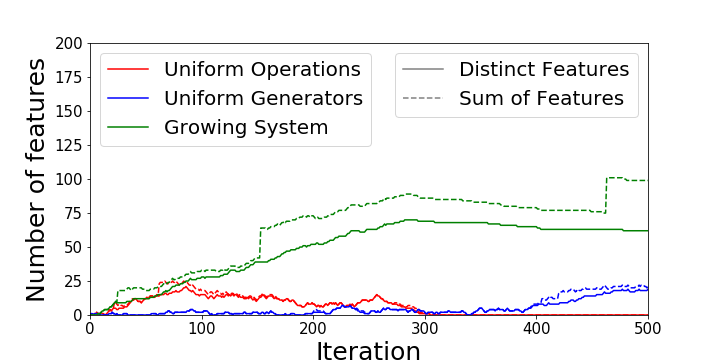}\label{fig:jsonNumFeat}}
\vspace{-.3cm}
\caption{Evolution of variability over first 500 iterations}
\vspace{-.5cm}
\label{fig:numFeat}
\end{figure}

Figure \ref{fig:numFeat} shows the evolution of the number of features as a measure of variability in the generated systems. We display both the number of distinct features and the sum of all features over all repositories. Due to the uniform distributions in \textit{Uniform Operators} and \textit{Generators}, %
added features are quickly removed again, resulting in constantly evolving low-variability systems. This is especially evident in the \textit{Uniform Generators} case in Figure \ref{fig:jsonNumFeat}. The growing system configuration on the other hand adds features more frequently than it removes them. It achieves a system with up to 76 transplanted features on the Calculator project. In most cases the variability starts to monotonically decrease at some point. This is due to no more features being available for transplantation in the donor systems.
In both cases this point is reached earlier in \textit{Uniform Operations} and \textit{Growing System} than in \textit{Uniform Generators}. This is reasonable, as the probability for selecting the feature addition generator is higher in these two configurations than in the latter one.
Note that the different configurations do not necessarily add the same features. 
In fact, while the difference in number of feature additions was limited to three features in the Calculator experiment, the \textit{Growing System} configuration added only 79 features compared to the 89 of both other parameterizations in the json-parser experiments.

\begin{figure}
\centering
\vspace{-.4cm}
\subfloat[Calculator]{\includegraphics[trim=0cm 0cm 0.3cm 0.5cm, clip, width=0.5\textwidth]{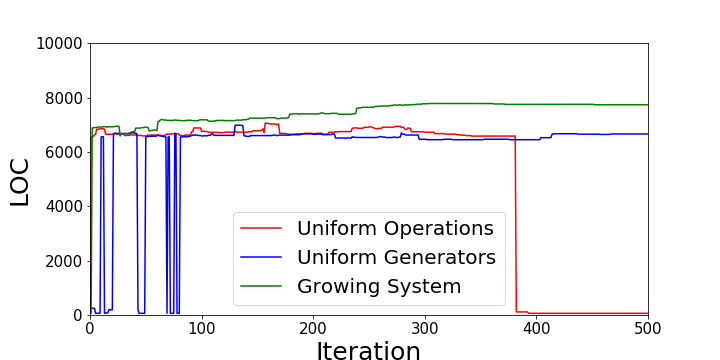}\label{fig:calcloc}}
\hfill
\vspace{-.4cm}
\subfloat[JSON-java]{\includegraphics[trim=0cm 0cm 0.3cm 0.5cm, clip, width=0.5\textwidth]{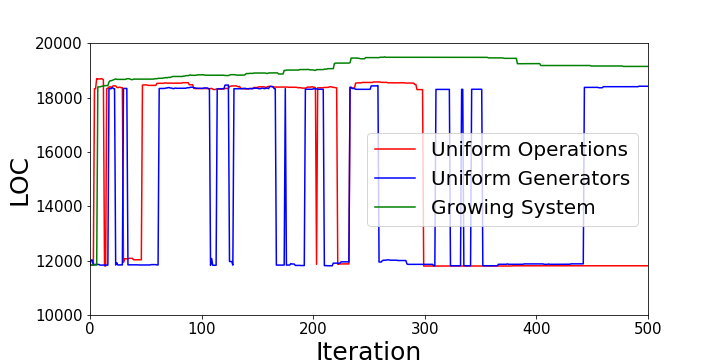}\label{fig:jsonloc}}
\vspace{-.3cm}
\caption{Evolution of system size over first 500 iterations}
\vspace{-.3cm}
\label{fig:loc}
\end{figure}

\looseness=-1
\Cref{fig:loc} shows the evolution of the size of the initial variant in lines of code (LoC).
The evolution histories for the two uniform probability configurations contain frequent large changes, while the growing system configuration results in a much more stable evolution. 
The reason for this is that a relatively large portion of the transplanted features has a large number of dependencies, which are added along with the testcase, if they are currently not present in the system. 
This can result in adding multiple thousands LoC.

\looseness=-1
Since we do not explicitly differentiate feature code from its dependencies, removing the only feature mapping to these dependencies results in a sharp cutback in code.
As discussed above, this happens frequently for uniform operations and generators, as they remove and add features with equal probability. 
An exception is the case of \textit{Uniform Generators} on the Calculator example: Five features were added early before cloning the repository, reducing volatility of the probability distribution by halving the likelihood of any operation targeting the repository.

\looseness=-1
On the other hand, the growing system configuration showcases a much more stable evolution with smaller additions and removals after the initial large code chunk, that is solidified by adding multiple features mapping to the same dependencies. In the JSON-java system we evolve from a single version with 11,837 LoC to 76,131 LoC over four variants.

\subsection{Performance (RQ4)}\label{sec:eval:rq2}

\begin{table}
\caption{Generation runtime for 500 iterations}
\label{tab:genruntime}
\vspace{-.4cm}
\begin{tabular}{lrr}
\toprule
& \multicolumn{1}{c}{calculator} & \multicolumn{1}{c}{JSON-java} \\ 
\midrule
\multicolumn{1}{l}{uniform operations} & 78 m 54 s & 202 m 48 s\\ 
\multicolumn{1}{l}{uniform generators} & 131 m 29 s & 239 m 11 s\\ 
\multicolumn{1}{l}{growing system} & 108 m 20 s & 125 m 32 s\\ 
\bottomrule
\end{tabular}
\vspace{-.5cm}
\end{table}

Table \ref{tab:genruntime} lists the runtimes of evolving both initial systems using three different configurations over 500 iterations. 
The main observation here is that the experiments evolving the smaller initial system finished quicker than the corresponding ones for the bigger systems.
While this seems expected at first, a closer look reveals that this discrepancy is mainly caused by the feature cloning generator. 
Calling the operation on the Calculator system took up to 17 minutes,
whilst taking up to 107 minutes on the json-parser. 
This is likely due to the substantially higher amount of traces that are added to the trace database when cloning repositories, that need to be checked for corresponding assets, when cloning features. %
We are aware that these results are by no means statistically significant and need to be backed up using evaluation data from a more exhaustive setup. 
However, these results indicate a performance bottleneck for the feature cloning generator, but provide otherwise encouraging evidence regarding the performance scalability of the other generators.

\section{Threats to Validity}
\parhead{External validity}
Our instantiation is partially specific to one programming language (Java) and build tool (Gradle). While instantiating it for other languages and tools would be desirable, \vpbench is language-independent and describes algorithms to implement the required language- and tool-specific parts. 

\looseness=-1
\parhead{Internal validity}
\Vpbench relies on various parameters.
Our case studies revealed that the configuration (choice of parameter values) strongly effects the plausibility of the generated version histories.
While we were able to find a configuration that leads to plausible outcomes, these parameters have to be tuned every time as soon as new generators are available.
Guiding the user in tuning the technique more systematically is a desirable direction for future work.

%% file: conclusion.tex
\section{Conclusion}
\looseness=-1
We presented a benchmark generation framework for evolving variant-rich software.
It simulates the evolution process of a variant-rich software system to generate a version history which can be included in a benchmark.
It relies on modular generators applying evolutionary changes---simple ones (e.g., mutating files, deleting features) and much more advanced ones (e.g., feature transplantation).

\looseness=-1
We believe the generated histories are useful for many uses cases, including evaluating \textit{feature identification and location}, \textit{change propagation}, \textit{refactoring}, and \textit{re-engineering} tooling. Feature location benchmarks are provided by utilizing the location meta-data. %
A benchmarking scenario for change propagation can be created by filtering the history for an asset- or feature propagation operation, which we intend to implement in the future. The problem (what to propagate) and the ground truth (how the propagation is done) can be extracted by taking the system versions before and after, respectively.
Another interesting use case is evaluating tools that recognize refactorings or re-engineerings in the generated version history. By making the intentions of those generated changes that are actually refactorings or re-engineerings explicit as meta-data, this allows build a reliable ground truth for such evaluations.
A different, but related use case would be to compare the generated refactorings or re-engineerings with actual refactorings and re-engineerings, suggested by automated tools. Explicitly defining and exploring all possible benchmarking scenarios is subject to future work, but a study on its own. We also plan to further enhance the generators, especially enhancing the naturalness of code\,\cite{hindle2016naturalness}.